\documentclass[]{pasj01}
\Received{2019/10/13}
\Accepted{2020/04/16}
\Published{2020/08/04}
 
\begin{document} 

\title{FLASHING: New high-velocity H$_2$O masers in IRAS 18286$-$0959}

%%% begin:list of authors
% Do NOT capitalize all letters in "textsc".
\author{Hiroshi  \textsc{Imai}\altaffilmark{1,2,3}, Yuri \textsc{Uno}\altaffilmark{3}, Daichi \textsc{Maeyama}\altaffilmark{4},  Ryosuke \textsc{Yamaguchi}\altaffilmark{4},  Kei \textsc{Amada}\altaffilmark{3}, Yuhki \textsc{Hamae}\altaffilmark{4}, Gabor \textsc{Orosz}\altaffilmark{5,6}, Jos\'e F.\ \textsc{G\'omez}\altaffilmark{7}, Daniel \textsc{Tafoya}\altaffilmark{8,9}, Lucero \textsc{Uscanga}\altaffilmark{10}, and Ross A. \textsc{Burns}\altaffilmark{8,11}}

\altaffiltext{1}{Amanogawa Galaxy Astronomy Research Center, Graduate School of Science and Engineering, Kagoshima University,  \\
1-21-35 Korimoto, Kagoshima 890-0065}
 \email{k3830453@kadai.jp}

\altaffiltext{2}{Center for General Education, Institute for Comprehensive Education, Kagoshima University,  \\ 
1-21-30 Korimoto, Kagoshima 890-0065}

\altaffiltext{3}{Department of Physics and Astronomy, Graduate School of Science and Engineering, Kagoshima University,  \\
1-21-35 Korimoto, Kagoshima 890-0065}

\altaffiltext{4}{Department of Physics and Astronomy, Faculty of Science, Kagoshima University,  \\
1-21-35 Korimoto, Kagoshima 890-0065}

\altaffiltext{5}{School of Natural Sciences, University of Tasmania, Private Bag 37, Hobart, Tasmania 7001, Australia}

\altaffiltext{6}{Xinjiang Astronomical Observatory, Chinese Academy of Sciences, 150 Science 1-Street, Urumqi, Xinjiang 830011, China}

\altaffiltext{7}{Instituto de Astrof\'isica de Andaluc\'{\i}a, CSIC, Glorieta de la Astronom\'{\i}a s/n, E-18008 Granada, Spain}

\altaffiltext{8}{National Astronomical Observatory of Japan, 2-21-1 Osawa, Mitaka, Tokyo, 181-8588, Japan}

\altaffiltext{9}{Department of Space, Earth and Environment, Chalmers University of Technology, Onsala Space Observatory, 439 92 Onsala, Sweden}

\altaffiltext{10}{Departamento de Astronom\'ia, Universidad de Guanajuato, A.P. 144, 36000 Guanajuato, Gto., Mexico}

\altaffiltext{11}{Korea Astronomy and Space Science Institute, 776 Daedeokdae-ro, Yuseong-gu, Daejeon 34055, Republic of Korea}

\KeyWords{masers --- stars: AGB and post-AGB --- stars: individuals(IRAS~18286$-$0959)} 

\maketitle

\begin{abstract}
We discovered new high-velocity components of H$_2$O maser emission in one of the ``water fountain" sources, IRAS~18286$-$0959, which has been monitored using the Nobeyama 45 m telescope in the new FLASHING 
(Finest Legacy Acquisitions of SiO- and H$_2$O-maser Ignitions by Nobeyama Generation) project since 2018 December. The maser spectra show new, extremely high expansion velocities ($>$200~km~s$^{-1}$ projected in the line of sight) components, some of which are located symmetrically in the spectrum with respect to the systemic velocity. They were also mapped with KaVA (KVN and VERA Combined Array) in 2019 March. We located some of these maser components closer to the central stellar system than other high velocity components (50--200~km~s$^{-1}$) that have been confirmed to be associated with the known bipolar outflow. The new components would flash in the fast collimated jet at a speed over 300~km~s$^{-1}$ (soon) after 2011 when they had not been detected. The fastest of the new components seem to indicate rapid deceleration in these spectra, however our present monitoring is still too sparse to unambiguously confirm it (up to 50~km~s$^{-1}$yr$^{-1}$) and too short to reveal their terminal expansion velocity, which will be equal to  the expansion velocity that has been observed ($v_{\rm exp}\sim$120~km~s$^{-1}$). Future occurrences of such extreme velocity components may provide a good opportunity to investigate possible recurrent outflow ignitions. Thus sculpture of the parental envelope will be traced by the dense gas that is entrained by the fast jet and exhibits spectacular distributions of the relatively stable maser features.
\end{abstract}

\section{Introduction}
\label{sec:introduction}
Our understanding of the final stellar evolution exhibiting copious and inhomogeneous mass loss and the subsequent formation of planetary nebulae (PNe) with a wide variety of their morphologies. It is considered that such transformation will happen during a relatively short period, from the end of the asymptotic giant branch (AGB) phase of stellar evolution to the beginning of the post-AGB phase, when an AGB stellar wind rapidly develops a circumstellar envelope (CSE) (e.g. \cite{lew01}). It has been confirmed that highly collimated, fast jets will be newly ejected from some of those stars (e.g. \cite{sah98}). These jets  shape the previously ejected CSE, so they may be ultimately responsible for the final morphologies of PNe.  However, two open questions remain: the launching mechanism and the timescale of such jet ejections. 

``Water fountain" (WF) sources have been identified as AGB or post-AGB stars hosting fast bipolar jets traced by H$_2$O maser emission \citep{ima07a}. The WF candidates were first confirmed in the maser spectra with extreme velocity widths much larger than those of typical 1612~MHz OH maser emission exhibiting clear double-horned spectral profiles (10--20~km~s$^{-1}$, e.g., \cite{lik92}, c.f., \cite{eng15}). There exist only 15 WFs confirmed to date \citep{gom17}, implying an extremely short timescale of the phase when such high velocity circumstellar H$_2$O masers are visible  ($<$100~yr, \cite{ima07a}) . Interferometric observations of these H$_2$O masers show a wide variety of the spatio-kinematical structures of the masers, reflecting different types of the central stellar systems driving these jets and/or different stages of the jet evolution. The diversity of the central star properties is indicated by the variety of the substructures in the maser spatio-kinematics such as multiple arcs, pairs of maser feature groups distributed point symmetrically, groups of maser features containing a large internal velocity dispersion ($>>$10~km~s$^{-1}$, e.g., \cite{ima02,cla09,ima13b,oro19}). 

The temporal evolution of each jet can be also monitored in the cases where its growth rate, namely the change rate of a total length of the maser jet on the sky, is consistent with the mean of the proper motion speeds of the individual maser features \citep{ima07b,cho15}. Recent observations of thermal molecular line and dust emission with the Atacama Large Millimeter-and-submillimeter Array (ALMA) have elucidated the complete spatio-morphologies of WF sources (e.g., \cite{sah17,taf19,taf20}). In some WFs, H$_2$O masers are suggested to trace the interaction of the fast jet with the parental CSE sculptured by the jet and resulting deceleration of the jet that leads entrained material along the jets. The jet deceleration is suggested by the spatio-kinematics of the H$_2$O masers themselves \citep{oro19}. However, other WFs show an opposite trend, with accelerating masers (e.g., \cite{gom15}), which could indicate a different nature of these WF phenomenon, maybe related to the evolutionary phase of the sources.

Taking into account the rarity of WFs, their possibly rapid evolution, and the possibility of the appearance of new H$_2$O masers in the known WFs and WF candidates, monitoring of these H$_2$O masers in single-dish observations with moderate cadence (1--8  weeks) is crucial for understanding the phenomena of WFs in a unified scheme. The possible deceleration of the jets may be proven by detecting H$_2$O masers at higher velocities with decreasing velocity drifts. A periodic behavior of the maser spectrum may indicate either periodic ejections of jets or pulsation of the central evolved star. In the latter case, synergistic monitoring of 1612~MHz OH masers and SiO (in addition to H$_2$O) masers will provide more concrete evidence for the stellar pulsation (e.g. \cite{her85}). The weakness of some velocity components of H$_2$O masers in WFs requires highly sensitive telescopes for systematic monitoring observations, and possibly an internationally coordinated program that can manage regular monitoring activities for the known WFs and WF candidates. The Nobeyama 45 m telescope is one of the best-suited telescopes for such monitoring observations thanks to its dynamic scheduling\footnote
{The dynamic scheduling as Back-up Program had been conducted until 2019 May.}, 
which provided good opportunities for high sensitivity H$_2$O maser observations. Moreover, the new quasi-optics enables us to simultaneously observe H$_2$O and SiO masers \citep{oka19}. Therefore we have monitored the WFs in the FLASHING (Finest Legacy Acquisitions of SiO and H$_2$O maser Ignitions by Nobeyama Generation) project since 2018 December. 

In this paper, we report the first scientific result of FLASHING, yielded for IRAS~18286$-$0959 (hereafter abbreviated as I18286). H$_2$O masers in I18286 have been mapped in several very long baseline interferometry (VLBI) observations using the Very Long Baseline Array (VLBA) and the VLBI Exploration of Radio Astrometry (VERA) (\cite{yun11}, Paper I; \cite{ima13a}, Paper II; \cite{ima13c}, Paper III). Paper I modeled the H$_2$O masers on a double helix pattern. Paper II determined the trigonometric parallax distance to I18286 to be  3.61$^{+0.63}_{-0.47}$~kpc. Paper III determined the locations of 1612~MHz OH masers and the low-velocity ($\sim$10~km~s$^{-1}$) components of the H$_2$O masers in the H$_2$O maser maps, both of which may be associated with a relic CSE in I18286. Before 2010, the H$_2$O masers in I18286 covered a velocity range of $-60\leq V_{\rm LSR}\leq 180$~km~s$^{-1}$ with respect to the local standard of rest (LSR). \citet{yun13} found new velocity components in 2011 around $V_{\rm LSR}\sim -90$ and 230~km~s$^{-1}$. This paper focuses on even newer, faster velocity components (surpassing the previously detected velocity range), found with our monitoring in FLASHING.

\section{Observations and data reduction}

\subsection{FLASHING observations} \label{sec:FLASHING}

While FLASHING is still active, in this paper we present the observed spectra towards I18286 until 2020 January 10. This dataset consist of eleven epochs: eight between 2018 December and 2019 May, and three additional epochs after 2019 November. Table \ref{tab:single-dish} gives the summary of the single-dish observations. We employed the new quasi-optics that enables us to observe simultaneously line emission at the frequency bands around 22.2 and 43.0~GHz \citep{oka19}, including the maser lines of H$_2$O $J_{K_{-}K_{+}}=6_{12}\rightarrow 5_{23}$ at the rest frequency of 22.235080~GHz and SiO $J=1\rightarrow 0$ ($v=3,\;2,\;1,$ and 0) in 42.5--43.4~GHz. This paper shows only the H$_2$O maser spectra because only this maser line was detected toward I18286. 

The aperture efficiency of the telescope is 0.65, yielding a flux-density conversion factor of 2.67~Jy~K$^{-1}$, when employing the new quasi-optics. The flux density scale was obtained from the information of the system noise temperature measured with the chopper-wheel method. The maser emission was received in both left- and right-hand circular polarization. Position switching was employed by alternating observations of a target maser and a blank sky position for an integration time of 30 s per scan. Three spectral windows were set to cover a bandwidth of 31.25~MHz in each window and centered at the mean velocity of the maser emission in the LSR frame ($\sim$65~km~s$^{-1}$) and those higher and lower by 30~MHz than the first one. The SAM45 spectrometers were used to obtain 4096 spectral channels per spectral window, but the final spectra had 2048 spectral channels after smoothing. The three spectral windows cover a velocity width of 1258~km~s$^{-1}$ in total with a velocity resolution of 0.42~km~s$^{-1}$. 

The data reduction was made using the Java NEWSTAR software\footnote
{See NEWSTAR home page: https://www.nro.nao.ac.jp/\textasciitilde nro45mrt/html/obs/newstar/index.html.}, 
in which integration and baseline subtraction were performed in order to obtain the final maser spectra. Dependent on system noise temperature and integration time, the 1-$\sigma$ noise level of the spectra ranged from 32 to 118 mJy. We identified detections of spectral components if their flux density was higher than the 5-$\sigma$ noise level (160--990 mJy). We then identified each spectral peak if it was composed of three or more detected components in consecutive spectral channels. In the case of multiple peaks, which are close in velocity and thus blended together, each peal was identified as an independent peak if its flux density was higher by 3-$\sigma$ than that of the spectral channel of the tail of a contiguous intersecting peak. 

\subsection{KaVA mapping observation} \label{sec:KaVA}
The H$_2$O masers in I18286 were also observed on 2019 March 4 for six hours with KaVA, an array composed of four telescopes of the Japanese VLBI Exploration of Radio Astrometry (VERA) plus three of the Korean VLBI Network (KVN). The fringe finder OT~081 and the delay calibrator J183005.9$+$061915 were also observed three times, every 25 min. These sources were observed in both of left- and right-hand circular polarization with the KVN telescopes, but only the former polarization data were valid. With VERA, only left-hand circular polarization was observed, but the dual-beam system allowed simultaneous observations of our target and the position reference source J183220.8$-$103511 (hereafter J1832). The received signals were filtered {into two base band channels (BBCs), each with a bandwidth of 128~MHz}, and recorded in 2-bit quantization, yielding a total recording rate of 1024Mbps. 

The data correlation was made using Daejeon Correlator in the Korea-Japan Correlation Center (KJCC). The recorded signals taken with both of the VERA's dual beams were correlated with those with the KVN telescopes. Two correlated data sets were provided from the two BBCs, including the maser source and J1832 respectively. 4096 spectral channels were contained for the BBC observing I18286, yielding a velocity channel spacing of 0.42~km~s$^{-1}$, while 128 spectral channels for other BBC observing J1832. Note that the latter data were invalid in the VERA--KVN baselines for the scans on J1832.

The data reduction was made using NRAO AIPS, which was handled using Python scripts and the ParselTongue package\footnote
{See ParselTonuge wiki: http://www.jive.nl/jivewiki/doku.php?id=parseltongue:parseltongue.}.  Standard a-priori calibration procedures were performed for calibrating visibility amplitudes using the information of the antenna gains and system noise temperatures. The delay calibration was performed using the scans on J183005.9$+$061915. Fringe fitting and self-calibration were performed using the data in a reference spectral channel at an LSR velocity of 139.5~km~s$^{-1}$, which included a bright and compact maser spot. Positional offsets of the maser spots given in this paper are referred to the position of this reference spot. Image cubes of the maser emission were created using the CLEAN deconvolution algorithm with a Gaussian synthesized beam of 2.1$\times$1.4 milliarcseconds (mas) at a position angle of $-18^{\circ}$, yielding a 1-$\sigma$ noise level of  $\sim$20~mJy in spectral channels without bright maser spots. Maser spots brighter than a 7-$\sigma$ noise level were identified in the individual spectral channels and grouped into isolated maser features, each of which includes maser spots located within 1--2~mas in position along consecutive spectral channels. We assume that the individual maser features correspond to physical features or gas clumps emitting the masers. 

The visibility data of J1832 observed only with VERA as well as the maser data were used for precise astrometry of the maser source.  These visibilities were calibrated using a more precise delay-tracking model. The data of the calibrator scans were used for instrumental delay calibration, including that for relative delay and phase offsets between VERA's dual beams. The fringe-fitting and self-calibration solutions obtained from the maser scans as mentioned above were applied to the J1832 data, yielding the relative position of J1832 with respect to that of the reference maser spot. However, the astrometric accuracy $\sim$0.4~mas is poorer than expected ($<$0.1~mas). This is a trade-off of the wide velocity coverage in the full KaVA array for the maser data instead of more accurate astrometry with VERA. The former observation setup yielded less sensitivity for J1832 in VERA, while the latter a limited velocity coverage of the maser source with the full KaVA. The self-calibration solutions taken from the maser data were also imperfect for astrometry due to insufficient performance of KVN whose accurate information of the station coordinates was still unavailable. 

\section{Results}

\subsection{Maser spectra}
\label{sec:spectra}
\begin{figure}[t]
\includegraphics[width=17.0cm]{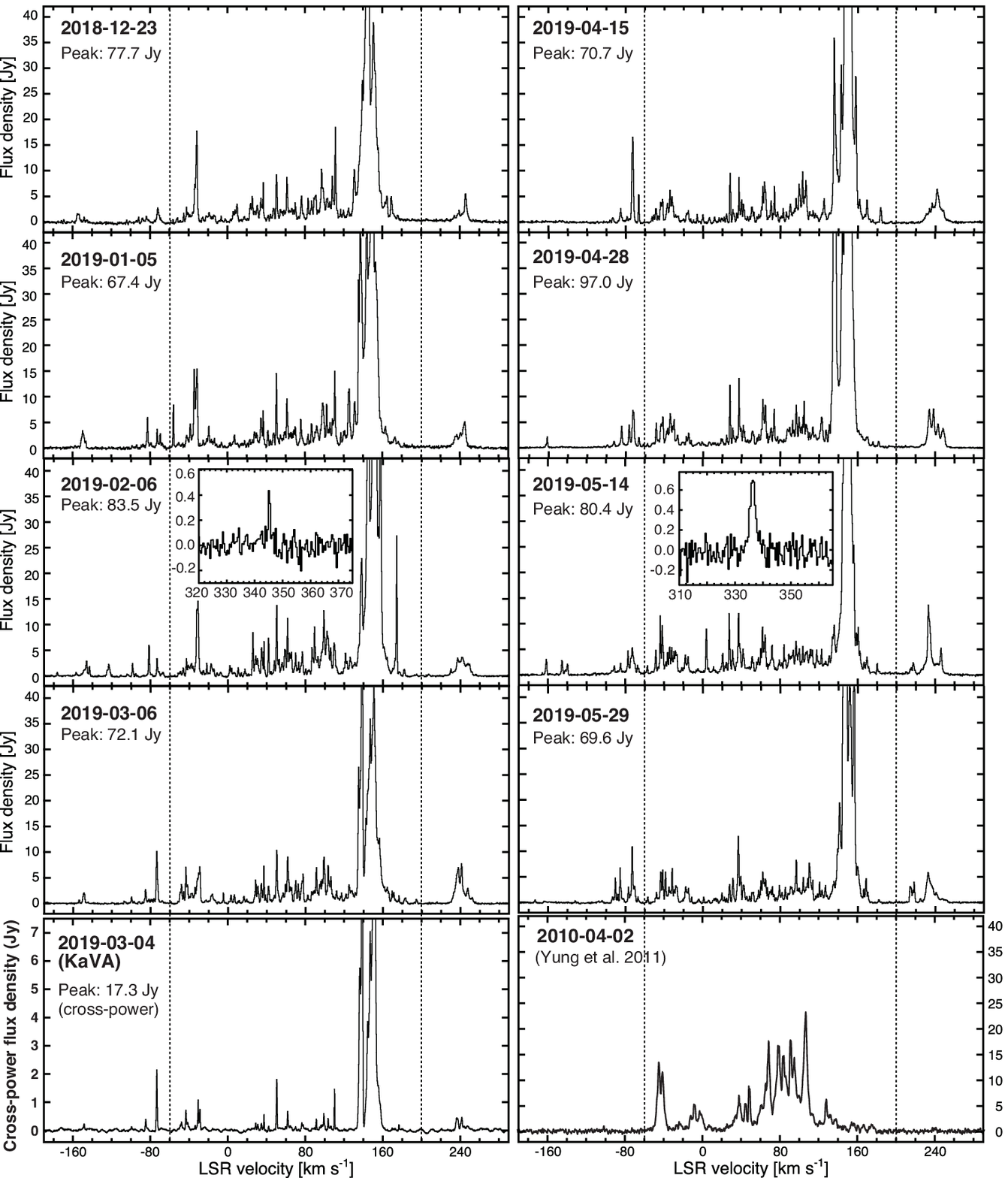} 
\caption{Time series of the H$_2$O maser spectra taken toward IRAS~18286$-$0959 using the Nobeyama 45 m telescope from 2018 December 23 to 2019 May 29. The insects of the spectra on 2019 February 6 and May 14 show the zoomed spectra of the further high velocity maser components detected at $V_{\rm LSR}\approx$336 and 345~km~s$^{-1}$. For comparison, the spectra taken on 2010 April 2 cited from Paper I is displayed in the right bottom panel with the same vertical scale as those in other 45 m spectra. The scalar-averaging cross-power spectrum taken in the KaVA observation is displayed in the left bottom panel. The two vertical dotted lines indicate the LSR velocities of $-$60 and 200~km~s$^{-1}$, over which the new high-velocity maser components appeared in the present observations.}\label{fig:spectra}
\end{figure}

Fig.\ \ref{fig:spectra} shows the spectra of H$_2$O masers in I18286 taken at the first eight epochs in FLASHING. They clearly show new components in $V_{\rm LSR}< -60$~km~s$^{-1}$ and $V_{\rm LSR}> 200$~km~s$^{-1}$, which had not been identified in the previous observations during 2006 April--2010 April (\cite{deg07}, Paper I). In our observations, we also find that the velocity components in the velocity range of $V_{\rm LSR}=$130--160~km~s$^{-1}$ are flaring compared to those seen in the previous observations by a factor of up to four.

The central velocity between the two extremely high velocity components present in all observing sessions (found around $V_{\rm LSR}\sim -$150 and 240~km~s$^{-1}$) is $V_{\rm LSR}$(center)$\simeq$45~km~s$^{-1}$, close to the velocity of the CO $J=3\rightarrow 2$ emission ($V_{\rm LSR}\simeq$65~km~s$^{-1}$, \cite{ima09}) and the velocity of the single 1612~MHz OH maser component  in I18286 ($V_{\rm LSR}$(OH)$\sim$40~km~s$^{-1}$, Paper III). In evolved stars, OH masers at 1612 MHz usually show a double-horned profile, with tracing an expansion of the CSE. Here we suggest that the central velocity of the CO emission traces the systemic stellar velocity  ($V_{\ast}\simeq$65~km~s$^{-1}$), while the velocity difference between the CO and OH peaks represents the expansion velocity of the relic envelope ($V_{\rm expansion}\simeq$25~km~s$^{-1}$). Furthermore, we found even higher velocity red-shifted components at $V_{\rm LSR}\simeq$345 and 336~km~s$^{-1}$ in the spectra on 2019 February 6 and May 14, respectively. As discussed later, it is unclear whether these represent the same gas clump (undergoing rapid deceleration, and a temporary maser quench between 2019 March and April), or they are independent clumps. In the latter case, each of the clumps would be short lived ($>$2 months). 

The maximum velocity separation between the fastest maser component (345~km~s$^{-1}$) and the systemic velocity is $\simeq$280~km~s$^{-1}$). This is much larger than the expansion velocity of the relic envelope mentioned above. Instead, we consider it as the lower limit of the projected jet velocity along the line of sight. This makes it one of the fastest water fountains, similar to IRAS~18113$-$2503 (hereafter I18113) \citep{gom11,oro19}. Adopting the distance to I18286 ($D=$3.6~kpc) and the maximum proper motion of the H$_2$O masers in the jet ($\sim$10~mas~yr$^{-1}$), the deprojected velocity of the jet would be $\sim$330~km~s$^{-1}$ at an inclination angle of $\sim$30$^{\circ}$ with respect to the line of sight. 

\begin{figure}[t]
\includegraphics[width=8.5cm]{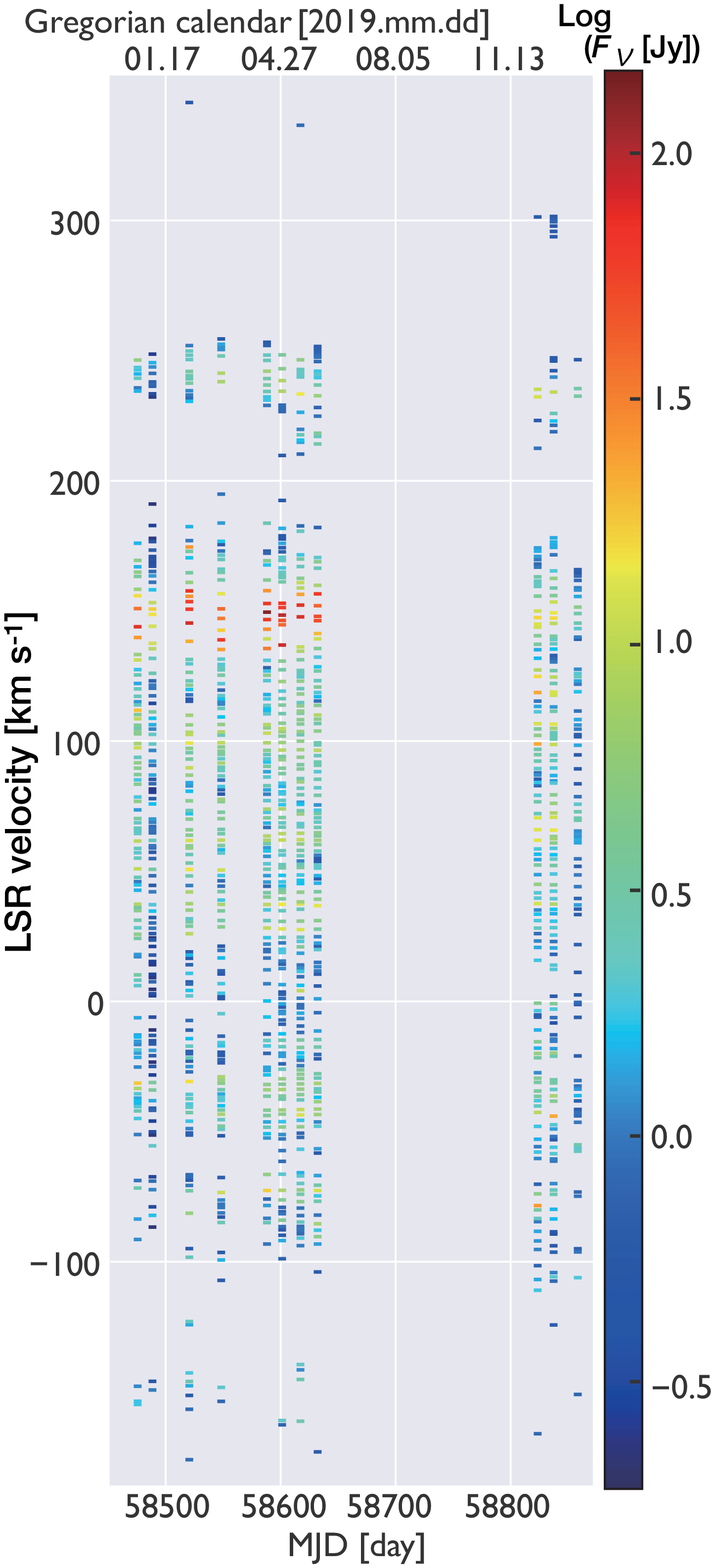} 
\caption{Velocity distributions of the spectral peaks ($>5\: \sigma$) of the IRAS~18286$-$0959 H$_2$O masers at the eleven epochs. A logarithm of the flux density of each spectral peak is indicated by the right-side color-scale bar.} 
\label{fig:peaks}
\end{figure}

Fig.\ \ref{fig:peaks} shows the time series of velocity distributions of the identified maser peaks. The time baseline of the FLASHING observations is still too short to trace any periodic variation in brightness or systematic drifts of the line-of-sight velocities of the maser peaks. It is also too short to relate the maser flare around $V_{\rm LSR}=$130--160~km~s$^{-1}$ with the new extremely high velocity components. One can see their large velocity drifts over one year, however it is unclear whether the components at $V_{\rm LSR}\simeq$345~km~s$^{-1}$ on 2019 February 6,  $\simeq$345 
~km~s$^{-1}$ on 2019 May 14, and $\sim$300~km~s$^{-1}$ in 2019 December--2020 January were really the same feature. Similarly, one also can note a velocity drift of the maser component at $V_{\rm LSR}\sim -$176~km~s$^{-1}$ on 2019 February 6, $\sim -173$~km~s$^{-1}$ on 2019 May 29, and  $\sim -$166~km~s$^{-1}$ on 2019 December 6. 

\subsection{Distribution of the maser features}
\label{sec:distribution}
\begin{figure}[t]
\includegraphics[width=8.5cm]{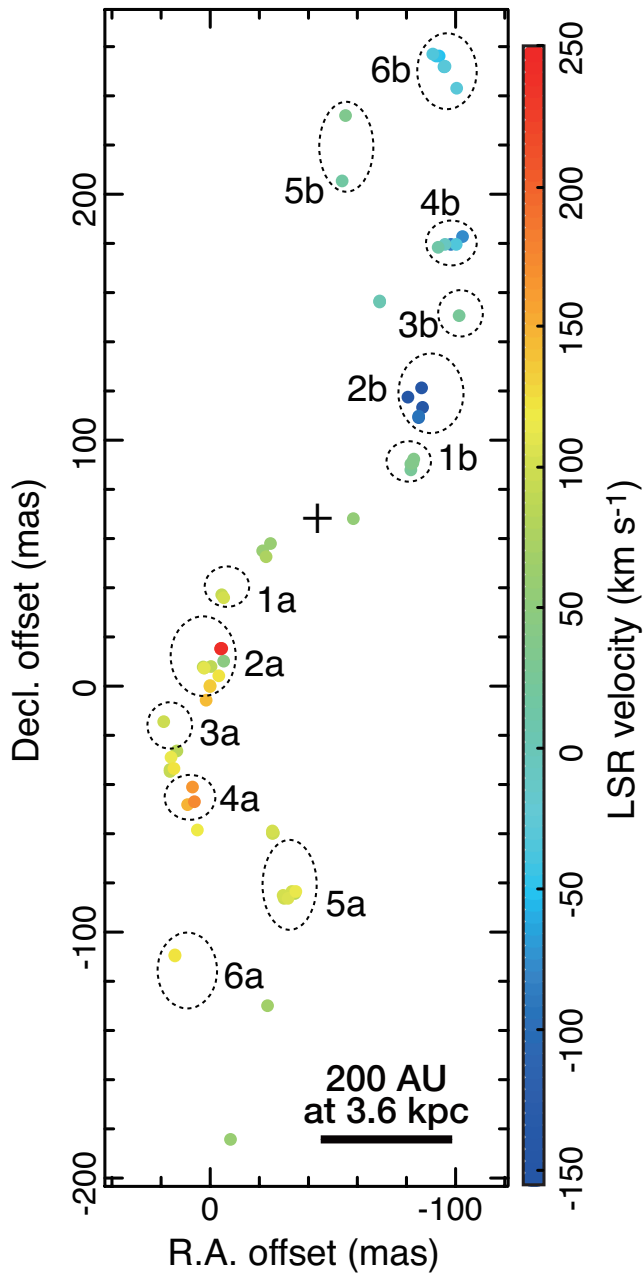}
\caption{Distribution of the H$_2$O maser features in IRAS~18286$-$0959 mapped with KaVA on 2019 March 4. The map origin is set to the maser spot whose emission was used for fringe-fitting and self-calibration as described in the main text. The color of each filled circle indicates the LSR velocity of the maser feature, whose scale is shown in the right-side color bar. The plus symbol denotes the estimated location of the stellar system driving the outflow. The labels of maser feature groups (from 1a-b to 6a-b) indicate point-symmetric locations of maser gas clumps with respect to the stellar system.}\label{fig:map}
\end{figure}

Table \ref{tab:features} gives the parameters of the I18286 H$_2$O maser features detected in the KaVA image cubes. The individual rows show, respectively, the LSR velocity at the intensity peak, the Right Ascension offset and its uncertainty, the declination offset and its uncertainty, the peak intensity, and the full velocity width of the feature. Fig.\ \ref{fig:map} shows the location of the H$_2$O maser components in I18286. These components are distributed in groups, tracing a point-symmetric structure with respect to the center, similar to W43A \citep{taf20}. In the figure, we label these symmetric groups with the same number, and with the letter a or b, depending whether they are located south or north of the center, respectively. The new extreme velocity components ($V_{\rm LSR}\approx -150$ and 240~km~s$^{-1}$) correspond to Group 2a and 2b. Group 2a is located close the the brightest and flared maser features (Group 4a).

\begin{figure}[t]
\includegraphics[width=8.0cm]{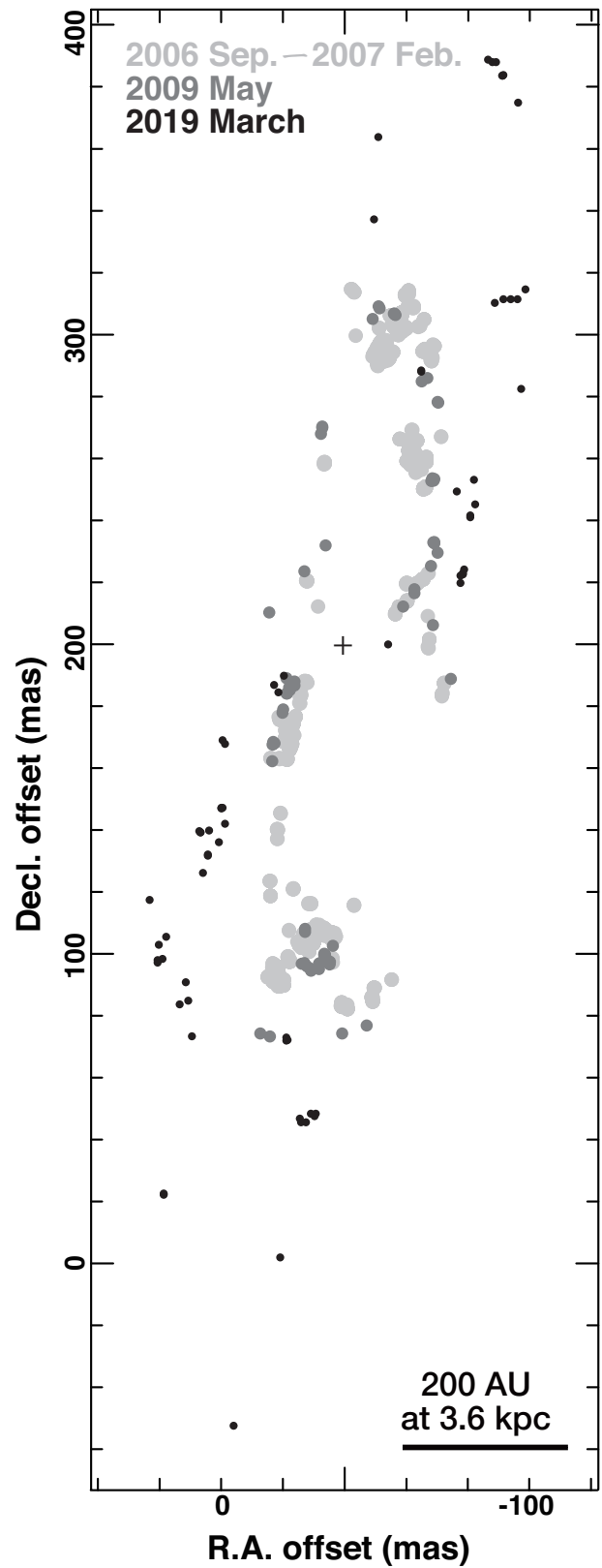} 
\caption{Comparison of the maser feature distributions in IRAS~18286$-$0959. Light grey, dark grey, and black filled circles indicate, respectively, the relative locations of maser features during 2006 September--2007 February, 2009 May, and 2019 March cited from Paper III, Paper II, and the present work. The plus symbol is as the same as that in Fig.\ref{fig:map}.}\label{fig:comparison}
\end{figure}

Fig.\  \ref{fig:comparison} shows the comparison among the maser feature distributions in 2006--2007, 2009, and 2019, and clarifies this secular variation. It is difficult to superimpose the maser maps in the common coordinate system precisely for comparison, because of the the uncertainty in the secular proper motion of the whole system. The astrometric information obtained in Paper II was composed of the Galactic rotation and the intrinsic motions of the individual maser features selected as position reference in the astrometry. Alternatively, map registration shown in Figure \ref{fig:comparison} has been produced by roughly referencing the possible location of the dynamical center of the outflow. 

Although stream lines of the gas hosting these masers might be bent by a magnetic force as suggested elsewhere (e.g. \cite{vle06}), we here suppose that the individual maser features are moving in straight lines along the bipolar outflow or jet driven by the stellar system. In this case, the stellar system may be located at the middle point between the new extreme velocity components, $(X,Y)\approx(-43,66)$[mas] from the reference feature used in self-calibration. Comparing with the maser map in 2008--2009 (Fig.\ 3 of Paper I), even without the accurate astrometric information, we note that the pair of the extreme velocity components in the present observation are located around the same regions as the highest velocity components in the previous observations ($V_{\rm LSR}\approx -50$ and 160~km~s$^{-1}$, Paper I, II, III) or even slightly closer to the stellar system. 

We also note that the alignment axis of the highest velocity components seems to rotate clockwise in this observation with respect to those in the previous epochs. This clockwise rotation of the maser alignment does not agree with the expectation from the previous kinematic model of two highly collimated precessing jets (Paper I), in which the precession is counter-clockwise (with a precessing angle and period of $\sim$28$^{\circ}$ and $\sim$56~yr, respectively). The spatio-kinematics of the outflow now seems to be  better explained by a model of wider-angle bipolar lobes from a single driving source as proposed for W~43A \citep{cho15}, in which H$_2$O masers are located along a bipolar cavity evacuated by a series of precessing collimated outflows. This scenario was also considered in Fig.\ 13 of Paper I, but was not favored there due to its inconsistency with the apparent systematic variation of the LSR velocities of the maser features along the outflow. However, \citet{taf19} and \citet{taf20} proposed a more plausible model of a bipolar outflow, which is composed of a faster collimated jet and slower lobes surrounding the jet. They suggest that H$_2$O masers are excited in the lobes formed in the material entrained by the jet. Although we see a very rapid expansion of the whole maser distribution by $\sim$140~mas in 13 yr, the expansion rate is consistent with the mean of the maser feature proper motions (8--10~mas~yr$^{-1}$, Paper I, II, and III). Thus we are likely watching the evolution of the I18286 jet traced by the H$_2$O masers (Section \ref{sec:discussion}). 

\begin{figure}[t]
\includegraphics[width=8.5cm]{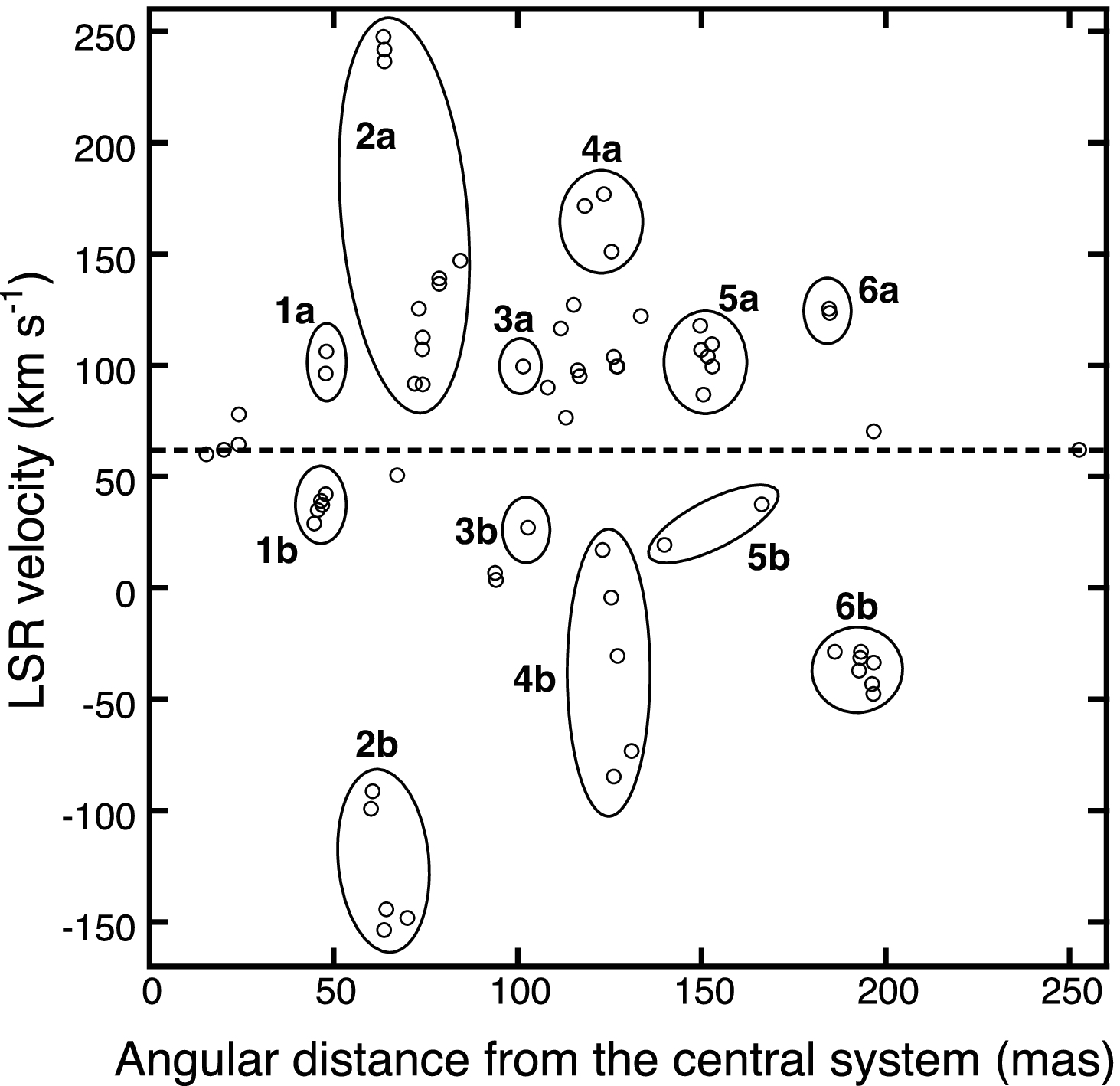} 
\caption{Distribution of the maser features in IRAS~18286$-$0959 in a velocity--distance diagram. The LSR velocity against the distance from the central stellar system, assumed to be at $(X, Y)=$($-$43, 66)[mas], is displayed for each maser feature. A horizontal dashed line shows the assumed systemic velocity of the central stellar system, $V_{\rm LSR}=$65~km~s$^{-1}$. The labels of maser feature groups (from 1a-b to 6a-b) are the same as those in Fig.\ \ref{fig:map}.}\label{fig:feature-distance}
\end{figure}

Fig.\ \ref{fig:feature-distance} shows the distribution of the maser features in a velocity--distance diagram, suggesting episodic gas ejections traced by the H$_2$O masers. Taking into account the spacing between the groups of maser components (20--30 mas) and the mean maser proper motions (8--10~mas~yr$^{-1}$), the time difference between the eruptions is roughly estimated to be 2--4~yr. The mean LSR velocity of the maser feature group pairs (1a--b, 3a--b, 5a--b) is derived to be $\sim$75~km~s$^{-1}$. Although the deviation of the mean velocity from the systemic velocity ($\Delta V \sim$10~km~s$^{-1}$) will provide a clue of the kinematics of the central stellar system, it should be examined in future thermal line mapping observations such as those made with ALMA. 

Unfortunately, the highest velocity components seen in our single-dish monitoring campaign ($V_{\rm LSR}\approx$336 and 345~km~s$^{-1}$) could not be detected in the KaVA session. On the other hand, thanks to the precise astrometry of KaVA, we determined the absolute coordinates of the maser feature containing the position-reference maser spot at $V_{\rm LSR}=$139.5~km~s$^{-1}$:  R.A.(J2000)$=18^{\rm h}31^{\rm m}22^{\rm s}$\hspace{-2pt}.93440$\pm 0^{\rm s}$\hspace{-2pt}.00002, decl.(J2000)$=-09^{\circ}57^{\prime}21^{\prime\prime}$\hspace{-2pt}.6654$\pm 0^{\rm \prime\prime}$\hspace{-2pt}.0003. This is useful for future precise registration between the maser emission and other spectral line maps.  

\section{Discussion}
\label{sec:discussion}
The previous single dish observations in 2006 April--2010 April (Paper I) did not detect the new extremely high velocity components ($V_{\rm LSR}< -60$~km~s$^{-1}$ and $V_{\rm LSR}> 200$~km~s$^{-1}$) reported in this paper, with a 5-$\sigma$ upper limit to 0.57~Jy, which suggests that these components would appear after 2010 April. We note that the velocity coverage of our VERA and VLBA observations (22 epochs in total during 2006 September--2009 September, Paper I, II and III) did not cover the velocity ranges including the new components, therefore we cannot ascertain whether such extreme components existed during those observations. One may wonder whether the material now showing extreme-velocity maser components was ejected much before their maser emission could be excited. However, this is unlikely. The material ejected at the expected initial speed ($>$300~km~s$^{-1}$) would interact within 1--2~yr of the ejection with the ambient gas that excites the maser features located within a few ten AU ($\sim$10 mas) of the stellar system (Paper III). The travel time of the ejected material to arrive at the present positions would be only 5~yr or shorter. 

The absence of the extreme components before 2010 (Paper I) should indicate the absence of the gas corresponding to these maser components at that time. \citet{yun13} found new velocity components in 2011 around $V_{\rm LSR}\sim -90$ and 230~km~s$^{-1}$. However, the absence of information of their locations makes it difficult for us to suggest that they were also associated with the same gas clumps as those hosting our discovered maser components. In fact, there is a relatively large shift ($>$10~km~s$^{-1}$) between the extreme velocity components in Paper I and those in this paper.

Previous observations could have missed the high-velocity components ($<\: -$60 and $>$200~km~s$^{-1}$) if their lifetime was short, as seen in the case of the components at 336 to 345~km~s$^{-1}$ and $-$178 to $-$144~km~s$^{-1}$ in this paper. However, the components around $-$90 to $-$70 and 230 to 250~km~s$^{-1}$ are actually relatively stable as seen in our monitoring observations, so we consider these components as a product of a new ejection after 2010 April. Here we propose that the stable high-velocity components seen at present would have formed soon after 2010 April with much higher velocities, like those seen in the unstable velocity components  observed in 2019 February and May (or the 336 and 345~km~s$^{-1}$ components). If this is the case, they could have already been quenched during the rapid deceleration of the jet, while the most extreme components in this paper would be excited by a new, more recent jet eruption. 

Here we note the colocation of the extreme velocity components in the southern side of the maser pair 2a and 2b with the components that have much lower velocities and flared flux densities in Group 4a (Fig.\ \ref{fig:map}). This is also consistent with the scenario suggested by \cite{taf20}, in which a newly jet is interacting with the ambient, slower material that is then accelerated as entrained material along the jet. However, any physical relationship between the extreme velocity and flared components, including the information of their relative locations along the line of sight, should be examined with further kinematic information.
 
On the other hand, as shown in Fig.\ \ref{fig:map}, we can see a point-symmetric pattern in the distributions of maser feature groups with respect to the central stellar system. This implies recurrent maser activity with a period of 2--4 yr (Section \ref{sec:distribution}) or longer together with recurrent gas eruptions. Taking into account the present size of the maser distribution ($\sim$260~mas in the largest distance from the central stellar system) and its growth rate (8--10~mas~yr$^{-1}$), the launching of the jet traced by H$_2$O is estimated to have happened in the year 1990$\pm$ 7 years. Here also note the existence of low velocity components ($25<V_{\rm LSR}<$105~km~s$^{-1}$) that are located closer to the central system than the new components, such as a pair of maser features (1a and b in Fig.\ \ref{fig:map}) and others possibly associated with the relic AGB envelope (Paper III). It is possible to consider that a pair of bipolar gas ejections may host not only the pair of the maser features mapped in this paper such as 2a and 2b, but also the fastest velocity components identified in single-dish spectra, but that are missing in this map. Similarly to the estimate described in Sect. \ref{sec:distribution} and as supposed above, one can consider that gas clumps hosting these most extreme velocity, short-lived  components were ejected within 2--4 years of the present day. 

Here we also consider another scenario for the I18286 jet, where a jet has been launched at a projected expansion velocity of $\sim$280~km~s$^{-1}$ as seen in the short-lived most extreme velocity components, which then decelerated to $\sim$110~km~s$^{-1}$ as seen in the stable components detected over the last 10 years due to drag by the ambient material in the CSE. This model is similar to that presented for I18113 by \citet{oro19}. Adopting the inclination angle of $\sim$30$^{\circ}$ these  projected expansion velocities correspond to $\sim$330~km~s$^{-1}$ and $\sim$130~km~s$^{-1}$, respectively. The jet velocity $v$ is now expressed as $v(t)=v_0 (1+v_0 k t)^{-1}$, where $v_0$ is the initial velocity of the jet,  $k$ the coefficient of the exponential decay of the expansion velocity as a function of the mass density of the CSE, the mass of the individual jet clumps, the drag coefficient of the jet clumps, and the cross-sectional area of the jet clumps. We adopt $v_0=$330~km~s$^{-1}$, $v(t)=$130~km~s$^{-1}$. When inserting $t\sim$30~yr as the dynamical age of I18286 (total length divided by a growth rate of the outflow estimated from maser locations and proper motions), $k$ is derived to be $\sim4.9 \times 10^{-17}$~cm$^{-1}$, which is well consistent with the value derived for IRAS~18113$-$2503 ($k\sim 6.4 \times 10^{-17}$~cm$^{-1}$, \cite{oro19}).

However, one should remember the absence of the fastest maser features 10 years ago and take into account the possibility that such components had already been decelerated to 
the velocity of the stable components ($\sim$110~km~s$^{-1}$) within 10 years. In this case, we get a much larger $k$ value, and a velocity drift over 10~km~s$^{-1}$yr$^{-1}$ should be observed and easily confirmed by weekly monitoring observations of the maser spectrum with a velocity resolution better than 1~km~s$^{-1}$. In our present work, we could not confirm such extremely fast velocity drifts due to the sparseness of the monitoring observations. On the other hand, we can see a general trend in which the whole maser distribution seems to exhibit accelerations from Group 1a to 6a and from 1b to 6b (Fig. \ref{fig:comparison}). Note that the distant maser features have an angular offset in the radial ejection direction from that of the closer and higher velocity components. A possible explanation for the distant features is that they are moving more slowly in a less dense region of the outer CSE and therefore, without the strong deceleration suggested for the high-velocity components. This is possible if we consider the precession of the jet. The first ejections have cleared up the CSE along the jet major axis, so that later ejections along that direction would travel more smoothly and with less deceleration than the highest-velocity ejections we see now along a different direction.

Paper I suggested the existence of twin jets, although it is unclear whether they are driven by a single stellar component or individual components in a stellar binary. The present results do not support that the twin jets are precessing anti-clockwise. This is evident when comparing the whole maser feature distributions over a decade as shown in Fig.\ \ref{fig:comparison}. Even with this rough superimposition of the maser maps as described in Sect.\ \ref{sec:distribution}, it is clear that some of the representative pattern of the maser features such as arcs have been moving radially from the dynamical center at a rate of $\sim$10~mas~yr$^{-1}$, as expected from the speed of the jet projected on the sky ($\gtrsim$150~km~s$^{-1}$, Paper I, II, II). The maser distribution seems to have been growing in the western and eastern sides in the north-west and south-east lobes, respectively, suggesting clockwise jet precession and lobe development.  

Nevertheless, it is noteworthy that the dynamical ages of the twin jets modeled in Paper I had a difference by $\sim$10~yr. Interestingly, the value of this time lag roughly corresponds to 2--3 cycles of the possibly repeating outbursts. Thus it is possible that some of the outbursts of a single jet could be further enhanced with a longer period as mentioned, so as to form arc-shaped structures of maser distribution as seen in I18113 \citep{oro19}. 

In I18286, even with a single jet, enhanced periodic outburst, if they exist, may form maser distributions that apparently look like twin jets as illustrated in Fig.\ 14 of Paper I. The model simulations by \citet{vel12} also reproduce multiple bipolar lobes produced by a precessing bipolar jet from a binary system. Although those models focused on planetary nebulae with nebular lobe lengths of $\sim$10$^{4}$~AU and binary orbital periods of $\sim$10~yr, one can consider a closer binary system for the case of I18286. We hope that morphological evolution of the H$_2$O masers will be further elucidated in the next decade, enabling us to compare it with that predicted by the models, in order to derive the physical parameters of the jet and the possible binary system. 

\section{Conclusions}
The spectroscopic monitoring in FLASHING and the VLBI follow-up imaging with KaVA of the H$_2$O masers in I18286 have revealed newly excited, extremely high velocity components in the H$_2$O masers and a corresponding axisymmetric ejections of gas in the water fountain outflow. In this paper, we discussed the possibility that the stellar system driving the outflow has developed its outflow lobes with a wide opening angle through periodic gas ejections with a clockwise precession of the outflow axis, rather than through highly collimated twin jets with an anti-clockwise axis precession. The new model explains the observed maser feature distributions biased to the down-stream sides of the bipolar lobes, namely the western and eastern sides of the northern and southern lobes, respectively. 

Further VLBI monitoring observations of the new components of H$_2$O masers will enable us to find their decelerations, suggested by \citet{taf19} and \citet{taf20}, as a result of interactions between the ambient material and the entrained material around the high-speed jet. Thus the combination of single-dish maser monitoring programs with targeted follow-up VLBI mapping observations provide an effective approach to investigating the ejection characteristics of water fountain jets/outflows. If the individual features in a variety of PN shapes are related to the individual events of such ejection behaviors seen in maser observations, this kind of approach will provide essential insights into the final stages of stellar evolution.

New ejections of the jet in I18286 should again be detected within several years if they are recurrent with time separations of 2--4 yr, as predicted by the spacing between maser groups. Since the size of the central binary system is expected be only a few AU (e.g., \cite{vel12}), which may correspond to a few mas or smaller, it is challenging to spatially resolve the system into a thermal collimated jet  from the driving star and an equatorial disk/torus driving the jet, even with state-of-the-art instruments such as ALMA. Therefore, the H$_2$O masers mapped with VLBI provide unique probes to elucidate the spatio-kinematical information and a possible periodic behavior of the central stellar system and the jet in I18286.

%%%%%%%%%%%%%%%%%%%%%%%%%%%%%%%%%%%%%%%

\begin{table}
  \tbl{Summary of the FLASHING observations toward IRAS~18286$-$0959.}{%
  \begin{tabular}{cccc}
      \hline
      Epoch & $T_{\rm sys}$\footnotemark[$*$]  & $t_{\rm int}$\footnotemark[$\dag$] & Noise\footnotemark[$\ddag$] \\ 
      (yy/mm/dd) &  (K) & (m) & (mJy) \\   \hline
18/12/23  &  $\sim$150 & 18 & 118 \\
19/01/05  &  $\sim$110 & 106 & 32 \\
19/02/06 & $\sim$150 & 47 & 77  \\
19/03/06 & $\sim$130 & 35 & 78  \\
19/04/15 & $\sim$190 & 58 & 114 \\ 
19/04/28 & $\sim$130 & 39 & 56 \\ 
19/05/14 & $\sim$170 & 37 & 108 \\
19/05/29 & $\sim$230 & 152 & 81 \\ 
19/12/06 & $\sim$120 & 28 & 77 \\
19/12/20 & $\sim$100 & 49 & 53 \\
20/01/10 & $\sim$110 & 23 &74 \\ 
\hline \hline
    \end{tabular}}\label{tab:single-dish}
\begin{tabnote}
\footnotemark[$*$] System noise temperature.  \\ 
\footnotemark[$\dag$] On-source integration time. \\
\footnotemark[$\ddag$] 1-$\sigma$ noise level of the spectrum. \\ 
%\footnotemark[$\S$]  ... \\ 
%\footnotemark[$\|$]  ... \\
%\footnotemark[$\sharp$]  ... \\  
%\footnotemark[$**$]  ... \\ 
%\footnotemark[$\dag\dag$]  ... \\ 
\end{tabnote}
\end{table}

\begin{longtable}{r@{ }r@{}r@{ }r@{}r@{}r@{}r} 
\caption{H$_2$O maser features detected with KaVA}
\label{tab:features}
\hline
 $V_{\rm LSR}$ & $X$ & $\sigma_X$ & $Y$ & $\sigma_Y$ & $I_{\rm peak}$  & $\Delta V$ \\
\hspace*{\fill} [km s$^{-1}$] & \hspace*{\fill} [mas] &\hspace*{\fill} [mas] & \hspace*{\fill} [mas] &
\hspace*{\fill} [mas] & \hspace*{\fill} [Jy beam$^{-1}$]  & \hspace*{\fill} [km s$^{-1}$] \\
 \endfirsthead
\hline
\endhead
\hline
\endfoot
\hline
\endlastfoot  
\hline
$  -153.57$&$   -80.64$& 0.25 &$   117.44$& 0.50 &        0.12 &  0.84 \\
$  -148.22$&$   -86.16$& 0.09 &$   121.15$& 0.10 &        0.72 &  3.16 \\
$  -144.32$&$   -86.59$& 0.13 &$   113.39$& 0.11 &        0.17 &  0.63 \\
$   -99.16$&$   -84.99$& 0.06 &$   109.08$& 0.09 &        0.55 &  1.26 \\
$   -91.20$&$   -85.06$& 0.05 &$   109.63$& 0.13 &        0.35 &  0.84 \\
$   -84.60$&$   -98.09$& 0.06 &$   179.46$& 0.09 &        1.18 &  2.32 \\
$   -73.14$&$  -102.67$& 0.02 &$   182.60$& 0.02 &        3.49 &  2.32 \\
$   -47.55$&$   -93.18$& 0.18 &$   256.19$& 0.11 &        0.80 &  2.95 \\
$   -43.08$&$   -91.94$& 0.19 &$   256.16$& 0.12 &        1.33 &  6.11 \\
$   -37.15$&$   -95.23$& 0.20 &$   251.61$& 0.10 &        0.26 &  1.47 \\
$   -33.51$&$   -90.69$& 0.07 &$   256.88$& 0.12 &        0.61 &  2.11 \\
$   -31.29$&$   -95.45$& 0.07 &$   251.87$& 0.09 &        0.56 &  0.63 \\
$   -30.42$&$  -100.15$& 0.05 &$   179.59$& 0.09 &        1.34 & 1.05 \\
$   -28.68$&$   -95.74$& 0.12 &$   252.00$& 0.10 &        1.11 &  1.69 \\
$   -28.65$&$  -100.54$& 0.11 &$   243.10$& 0.11 &        0.56 &  2.53 \\
$    -4.21$&$   -95.82$& 0.10 &$   179.80$& 0.05 &        0.36 &  0.63 \\
$     3.59$&$   -69.07$& 0.08 &$   156.41$& 0.06 &        0.45 &  1.90 \\
$     6.87$&$   -68.95$& 0.08 &$   156.23$& 0.14 &        0.24 &  1.26 \\
$    17.07$&$   -92.67$& 1.21 &$   178.60$& 0.35 &        0.37 & 2.74 \\
$    19.39$&$   -53.75$& 0.09 &$   205.44$& 0.11 &        0.23 &  0.63 \\
$    27.17$&$  -101.41$& 0.08 &$   150.62$& 0.18 &        0.21 &  0.63 \\
$    28.93$&$   -81.87$& 0.06 &$    88.11$& 0.36 &        0.99 &  4.00 \\
$    34.97$&$   -81.64$& 0.47 &$    90.45$& 0.66 &        0.80 &  1.26 \\
$    37.29$&$   -82.59$& 0.12 &$    91.21$& 0.23 &        0.43 &  1.47 \\
$    37.52$&$   -55.05$& 0.05 &$   231.91$& 0.08 &        1.35 & 0.84 \\
$    39.19$&$   -82.44$& 0.13 &$    90.83$& 0.09 &        0.32 &  0.63 \\
$    42.17$&$   -82.94$& 0.05 &$    92.29$& 0.10 &        0.79 &  1.69 \\
$    50.75$&$    -5.43$& 0.02 &$    10.23$& 0.04 &        2.83 &  2.11 \\
$    60.02$&$   -58.33$& 0.05 &$    68.19$& 0.11 &        0.54 &  0.84 \\
$    62.07$&$   -24.43$& 0.09 &$    58.00$& 0.17 &        1.14 &  1.26 \\
$    62.09$&$    -8.24$& 0.31 &$  -184.19$& 0.18 &        0.59 &  0.84 \\
$    64.61$&$   -21.45$& 0.11 &$    55.00$& 0.21 &        0.48 &  2.11 \\
$    70.37$&$   -23.38$& 0.30 &$  -129.79$& 0.10 &        0.37 &  0.63 \\
$    76.65$&$    11.50$& 0.11 &$   -33.09$& 0.15 &        0.70 & 2.53 \\
$    78.05$&$   -22.78$& 0.10 &$    52.43$& 0.22 &        0.56 &  1.27 \\
$    87.00$&$   -34.51$& 0.20 &$   -84.24$& 0.06 &        0.35 &  1.05 \\
$    90.09$&$    13.58$& 1.15 &$   -26.22$& 0.16 &        0.41 &  0.84 \\
$    91.52$&$     2.93$& 0.13 &$     7.71$& 0.05 &        1.03 &  1.68 \\
$    91.76$&$    -0.28$& 0.42 &$     7.94$& 0.26 &        0.79 &  1.68 \\
$    95.11$&$    16.40$& 0.11 &$   -34.72$& 0.11 &        0.49 &  1.05 \\
$    96.26$&$    -4.82$& 0.13 &$    37.17$& 0.06 &        0.55 &  0.84 \\
$    97.78$&$    16.29$& 0.05 &$   -34.13$& 0.07 &        0.77 &  2.32 \\
$    99.41$&$   -25.51$& 0.31 &$   -59.97$& 0.09 &        0.59 &  1.05 \\
$    99.46$&$   -25.84$& 0.37 &$   -59.79$& 0.15 &        0.60 &  1.26 \\
$    99.46$&$   -30.18$& 0.16 &$   -86.37$& 0.06 &        1.34 &  2.95 \\
$    99.47$&$    18.93$& 0.61 &$   -14.49$& 0.21 &        0.55 &  2.53 \\
$   103.85$&$   -25.26$& 0.51 &$   -58.82$& 0.31 &        0.42 &  2.74 \\
$   103.87$&$   -29.62$& 0.31 &$   -85.11$& 0.49 &        1.40 & 2.53 \\
$   106.33$&$    -5.66$& 0.07 &$    35.81$& 0.07 &        0.73 &  2.32 \\
$   106.93$&$   -33.42$& 0.05 &$   -83.59$& 0.04 &        0.86 &  1.27 \\
$   107.31$&$     2.58$& 0.15 &$     7.48$& 0.14 &        0.35 &  2.32 \\
$   109.50$&$   -31.68$& 0.10 &$   -86.39$& 0.15 &        0.24 &  0.63 \\
$   112.58$&$     2.45$& 0.08 &$     7.32$& 0.12 &        0.52 &  1.47 \\
$   116.56$&$    15.86$& 0.16 &$   -28.99$& 0.11 &        0.51 & 4.64 \\
$   117.88$&$   -34.86$& 0.08 &$   -83.40$& 0.06 &        0.59 &  1.05 \\
$   122.20$&$     5.33$& 0.08 &$   -58.43$& 0.11 &        0.72 &  2.11 \\
$   123.68$&$    14.24$& 0.10 &$  -109.73$& 0.15 &        0.27 &  1.26 \\
$   125.38$&$    14.18$& 0.12 &$  -109.51$& 0.11 &        0.36 &  0.84 \\
$   125.36$&$    -3.66$& 0.08 &$     4.29$& 0.17 &        0.24 &  1.47 \\
$   127.19$&$    14.84$& 0.12 &$   -33.61$& 0.09 &        0.33 &  1.05 \\
$   136.59$&$     0.06$& 0.02 &$     0.04$& 0.09 &       11.96 &  6.32 \\
$   139.05$&$     0.01$& 0.02 &$     0.00$& 0.02 &       28.00 &  3.58 \\
$   147.08$&$     1.74$& 0.10 &$    -5.64$& 0.25 &        5.59 &  6.74 \\
$   151.10$&$     9.03$& 0.19 &$   -48.21$& 0.30 &       23.69 & 24.02 \\
$   171.52$&$     7.26$& 0.06 &$   -41.01$& 0.09 &        0.47 &  0.84 \\
$   177.03$&$     6.37$& 0.11 &$   -47.14$& 0.05 &        0.88 &  1.68 \\
$   236.62$&$    -4.37$& 0.08 &$    15.23$& 0.04 &        1.28 &  7.16 \\
$   241.89$&$    -4.43$& 0.11 &$    15.11$& 0.06 &        1.38 &  6.32 \\
$   247.53$&$    -4.84$& 0.19 &$    15.25$& 0.06 &        0.84 &  7.80 \\
\end{longtable}

%%%%%%%%%%%%%%%%%%%%%%%%%%%%%%%%%%%%%%%

\begin{ack}
The Nobeyama 45-m radio telescope and VERA are operated by Nobeyama Radio Observatory and Mizusawa VLBI Observatory, respectively, branches of National Astronomical Observatory of Japan (NAOJ), National Institutes of Natural Sciences. KVN in KaVA and the Daejeon Correlator in KJCC are operated by Korea Astronomy and Space Science Institute (KASI). Our data analysis was in part carried out on the common use data analysis computer system at the Astronomy Data Center, ADC, of NAOJ. HI and GO are supported by the MEXT KAKENHI program (16H02167). JFG is partially supported by MINECO (Spain) grant AYA2017-84390-C2-R (co-funded by FEDER) and by the State Agency for Research of the Spanish MCIU through the ``Center of Excellence Severo Ochoa" award for the Instituto de Astrof\'isica de Andaluc\'{\i}a (SEV-2017-0709). HI and JFG were supported by the Invitation Program for Foreign Researchers of the Japan Society for Promotion of Science (JSPS grant S14128) and i-LINK+2019 Programme in IAA/CSIC. GO was supported by the Australian Research Council Discovery project DP180101061 of the Australian government, and the grants of CAS LCWR 2018-XBQNXZ-B-021 and National Key R\&D Program of China 2018YFA0404602. DT was supported by the ERC consolidator grant 614264. RB acknowledges support through the EACOA Fellowship from the East Asian Core Observatories Association. LU acknowledges support from Convocatoria de Apoyo a la Investigaci\'on Cient\'if||\'ica 2020 of the University of Guanajuato.
\end{ack}

%----------------------------------------------------------------------------------------
%	BIBLIOGRAPHY
%----------------------------------------------------------------------------------------

\end{document}